\documentclass[aps,prx,reprint,superscriptaddress, amsmath, amssymb]{revtex4-2}

\usepackage{graphicx}
\usepackage{dcolumn}
\usepackage{bm}

\begin{document}

\preprint{APS/123-QED}

\title{\textbf{An Atomic Interface for High-Dimensional Temporal Mode Quantum Networks} 
}%

\author{Shicheng Zhang}
\thanks{These authors contributed equally to this work.}
\affiliation{Department of Physics, Imperial College London, London, SW7 2AZ, United Kingdom.}

\author{Aonan Zhang}%
\thanks{These authors contributed equally to this work.}
\affiliation{Department of Physics, Imperial College London, London, SW7 2AZ, United Kingdom.}
\affiliation{Department of Physics, University of Oxford, Oxford, OX1 3PU, United Kingdom}

\author{Ilse Maillette de Buy Wenniger}
\affiliation{Department of Physics, Imperial College London, London, SW7 2AZ, United Kingdom.}

\author{Paul M. Burdekin}
\affiliation{Department of Physics, Imperial College London, London, SW7 2AZ, United Kingdom.}

\author{Jerzy Szuniewicz}
\affiliation{Faculty of Physics, University of Warsaw, Warszawa, 02-093, Poland.}

\author{Steven Sagona-Stophel}
\affiliation{Department of Physics, Imperial College London, London, SW7 2AZ, United Kingdom.}

\author{Sarah E. Thomas}%
\email{sarah.thomas@eng.ox.ac.uk}
\affiliation{Department of Physics, Imperial College London, London, SW7 2AZ, United Kingdom.}
\affiliation{Department of Engineering Science, University of Oxford, Oxford, OX1 3PJ, United Kingdom}

\author{Ian A. Walmsley}
\affiliation{Department of Physics, Imperial College London, London, SW7 2AZ, United Kingdom.}

\date{\today}

\begin{abstract}

\noindent

Temporal modes of photons are a promising encoding scheme for high-dimensional quantum networks due to their high channel capacity and fiber compatibility. However, realizing their full potential requires devices capable of synchronizing, processing and interfacing these modes across photonic and atomic bandwidths. In this work, we demonstrate a programmable high-dimensional temporal mode processor using a Raman quantum memory in warm cesium vapor. We exploit the single-mode nature of the Raman interaction kernel, dynamically shaping the control field to synthesize a tunable coherent filter that selectively addresses specific temporal waveforms. This mechanism enables on-demand storage, filtering, and conversion, providing a coherent interface between MHz- and GHz-bandwidth modes. We validate the platform’s selectivity across a basis of 30 orthogonal Hermite-Gaussian modes and certify high-fidelity quantum operation via 5-dimensional process tomography. By combining deterministic mode conversion with bidirectional bandwidth interfacing, we establish the Raman memory as a critical active node for scalable quantum information processing.


\end{abstract}

\maketitle


\section{Introduction}

The realization of a global quantum internet as a distributed network capable of secure quantum key distribution~\cite{Cao_2022,Mehic_2020,Mamann_2025}, distributed quantum computing~\cite{Oh_2023,Main_2025}, and enhanced sensing~\cite{Degen_2017,Pirandola_2018,Ye_2024} relies on the scalable transmission of quantum information between remote nodes. Photons serve as the ideal carriers for this task due to their minimal decoherence during propagation. However, common architectures based on the two-dimensional polarization encoding face fundamental constraints: they are intrinsically limited to one bit of information per photon and possess lower error thresholds for security than high-dimensional systems. To overcome these bounds, high-dimensional encoding offers significant advantages, including increased channel capacity, improved resilience to noise in quantum key distribution~\cite{Cozzolino_2019}, and enhanced computational power for quantum simulation and processing~\cite{Wang_2018}.

Among the available degrees of freedom for high-dimensional encoding, the temporal modes (TMs) of light represent the most promising candidate for fiber-integrated networks~\cite{Brecht_2015,Raymer_2020}. Defined as field-orthogonal wavepackets within a single spatial mode~\cite{Brecht_2015}, TMs are immune to the spatial mode mixing that severely limits the fiber transmission of spatial states~\cite{Cao_2023}. Moreover, TMs offer distinct advantages over alternative temporal schemes such as time-bin encoding~\cite{Brougham_2016,Cozzolino_2019}. While bin-based encodings require scaling physical resources like timing resolution linearly with dimension, TMs allow for dense information packing within a fixed time window. Furthermore, TMs serve as the natural eigenbasis of parametric down-conversion sources, which inherently generate high-dimensional temporal entanglement~\cite{Graffitti_2020,Huo_2020}. Yet the widespread adoption of TM encodings is currently bottlenecked by the lack of efficient and programmable hardware capable of manipulating them. Because these modes overlap simultaneously in time and space,  separating them requires a time-non-stationary interaction that functions as a time-dependent beam splitter to selectively address specific temporal waveforms~\cite{Brecht_2015}.


To date, coherent TM manipulation has been pioneered using nonlinear optical processes like the quantum pulse gate~\cite{Ansari_2017,Shahverdi_2017,Allgaier_2017,Reddy_2018,Serino_2023,Serino_2025} and electro-optic time lenses~\cite{Karpinski_2017,Ashby_2020,Joshi_2022}. While these broadband architectures (tens-of-GHz to THz) excel at high-capacity transmission, they suffer from a severe spectral mismatch with the narrowband (MHz) atomic nodes, and lack the optical buffering required to synchronize asynchronous network events. Conversely, matter-based platforms, such as Raman or Autler-Townes memories~\cite{Fisher_2016,Saglamyurek_2018}, naturally offer integrated storage and bandwidth control, yet experimentally scaling them beyond simple Gaussian profiles to achieve programmable, high-dimensional TM control remains a significant challenge. Consequently, a unified interface capable of simultaneously storing, filtering, and converting complex photonic modes to bridge the gap between high-speed optical links and narrowband quantum processors is still missing for hybrid quantum networks.

\begin{figure*}
\includegraphics[width=0.9\linewidth]{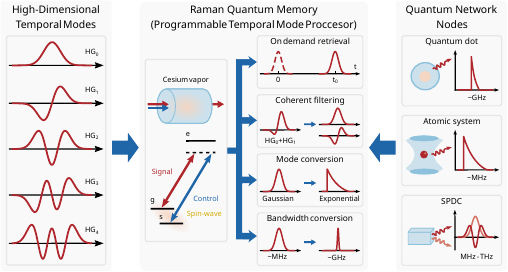}
\caption{\textbf{Conceptual overview of the Raman quantum memory as a versatile high-dimensional temporal mode processor and network interface.} The Raman quantum memory is capable of processing high-dimensional temporal modes, such as orthogonal Hermite-Gaussian wavepackets, by mapping them into and out of an atomic spin-wave via a programmable two-photon transition. By tailoring the classical control field, the memory performs on-demand retrieval, coherent filtering, as well as deterministic mode and bandwidth conversion. This versatility enables direct interfacing between disparate quantum network nodes, bridging the bandwidth gap between broadband emitters (e.g., GHz-bandwidth quantum dots and parametric down-conversion sources) and narrowband matter-based systems (MHz-bandwidth atomic nodes).}
\label{fig:Scheme}
\end{figure*}

In this work, we demonstrate that a Raman quantum memory in warm cesium vapor functions as a versatile, high-dimensional temporal mode processor. By independently shaping the write and read control fields, we can deterministically store, filter, and convert quantum states on demand. As conceptually illustrated in Fig.~\ref{fig:Scheme}, this dynamic programmability enables the memory to bridge the critical MHz-GHz bandwidth regime, establishing a universal interface between diverse quantum network nodes. We experimentally validate this capability by characterizing the storage selectivity across a basis of 30 orthogonal Hermite-Gaussian (HG) modes and strictly certifying the quantum performance via a 5-dimensional process tomography, which yields a reconstruction fidelity of $0.943 \pm 0.015$ relative to an ideal single-mode filter. Furthermore, we utilize the flexibility of the interaction to demonstrate deterministic mode conversion alongside bandwidth compression and expansion, establishing the Raman platform as a robust, active node for scalable high-dimensional quantum networks.

\section{Single Mode Raman quantum memory}

To understand the capability of the Raman memory as a programmable temporal mode processor, we introduce the dynamics of the light-matter interaction. The operational principle relies on the coherent mapping of a photonic field onto a stationary atomic coherence via an off-resonant two-photon interaction. The system comprises an ensemble of atoms in a $\Lambda$-type energy level configuration, where a weak signal field $\mathcal{E}$ and a strong control field $\Omega$ couple two hyperfine ground states, $|g\rangle$ and $|s\rangle$, to a virtual excited state $|e\rangle$ with a large single-photon detuning $\Delta$. This configuration facilitates a stimulated Raman scattering process that maps the input signal onto a long-lived collective spin-wave excitation $\mathcal{B}$ distributed across the ensemble. Subsequently, the application of a second control pulse triggers the reverse process, converting the stored spin-wave back into an optical field.

The ideal, noise-free Raman memory is governed by
a set of linear equations describing the evolution of the signal mode $\mathcal{E}$ and the spin-wave mode $\mathcal{B}$, driven by the control Rabi frequency $\Omega$~\cite{Nunn_2007}.
\begin{eqnarray}
\partial_z \mathcal{E} & = & + i  \sqrt{\frac{d \gamma}{L}} \frac{\Omega}{\Gamma} \mathcal{B} -\frac{d \gamma }{L}\frac{1}{\Gamma} \mathcal{E}; \label{eqn:IdealRaman_S} \\
\partial_\tau \mathcal{B} & = & -i \sqrt{\frac{d \gamma}{L}} \frac{\Omega^* }{\Gamma}  \mathcal{E} - \frac{|\Omega|^2}{\Gamma}  \mathcal{B}. \label{eqn:IdealRaman_B} 
\end{eqnarray}
Here, $d$ represents the optical depth of the atomic ensemble of length $L$, and $\Gamma = \gamma + i \Delta $ is the complex detuning, where $\gamma$ is the natural linewidth of the atomic transition.  
These equations describe a beam-splitter-like interaction where the ``reflectivity"—the coupling strength between the light and matter modes—is time-dependent and determined by the instantaneous control field $\Omega$.

Because the equations of motion are linear, the storage process that maps an input signal field $\mathcal{E}_{in}(\tau)$ to a stored spin-wave $\mathcal{B}_{stor}(z)$ can be described as,
\begin{equation}    
    \mathcal{B}_{stor}(z) = \int_{-\infty}^{\infty} K_1(z,\tau) \mathcal{E}_{in}(\tau) d\tau.
\label{eq:storage}
\end{equation}
The kernel $K_1(z,\tau)$, often referred to as the memory Green’s function, encapsulates the physics of the storage process. The storage efficiency is then defined as the ratio of the stored spin-wave excitation number to the input photon number: 
\begin{equation} 
\eta_{\text{stor}} = \frac{N_{\text{stor}}}{N_{\text{in}}} = \frac{\int_0^L dz \langle \mathcal{B}_{\text{stor}}^\dag(z)\mathcal{B}_{\text{stor}}(z) \rangle}{\int_{-\infty}^{+\infty } d\tau \langle \mathcal{E}_{\text{in}}^\dag(\tau)\mathcal{E}_{\text{in}}(\tau) \rangle}.
\label{eq:storage_eff} 
\end{equation}

To quantify the mode selectivity of this interaction, we perform a singular value decomposition of the storage kernel~\cite{Nunn_2008}, decomposing it into a sum of separable modes:
\begin{equation}    
    K_1(z,\tau) = \sum_{k} \lambda_{k} \psi_{k}(z) \phi_{k}^{*}(\tau).
\end{equation}
In this decomposition, $\{\phi_{k}(\tau)\}$ forms an orthonormal basis of temporal modes for the input light, and $\{\psi_{k}(z)\}$ forms the corresponding basis for the stored spin-wave. The singular values $\lambda_k$ quantify the storage efficiency $\eta_k = \lambda_k^2$ for each mode. In the low-coupling regime of Raman memory, the distribution of singular values is sharply peaked. The first singular value $\lambda_1$ is typically much larger than all higher-order values ($\lambda_1 \gg \lambda_{k>1}$), indicating that the memory effectively interacts with only a single temporal mode, $\phi_1(\tau)$. Consequently, the device functions as a temporal mode filter: it efficiently stores the mode defined by the control field while remaining transparent to all orthogonal modes. By tailoring the control field $\Omega(\tau)$, we can program the memory to select any arbitrary temporal mode from a high-dimensional superposition.

The retrieval process is the time-reversed dual of storage. A read control field $\Omega_{read}(\tau)$ maps the stored spatial spin-wave $\mathcal{B}$ back onto an output optical field $\mathcal{E}_{out}(\tau)$. The dynamics are governed by a retrieval kernel $K_2(\tau,z)$:
\begin{equation}    
    \mathcal{E}_{out}(t) = \int_{0}^{L} K_2(\tau,z) \mathcal{B}_{stor}(z) dz.
\end{equation}
This relationship reveals the mechanism for mode conversion. The temporal shape of the retrieved photon is not fixed by the stored spin-wave alone; rather, it is the result of the convolution between the stored spatial coherence and the read control field. Because the write and read processes are temporally distinct, the control fields can be shaped independently. By adjusting the temporal shape of the read control field, the retrieved mode can be deterministically modified in both mode number and temporal bandwidth, enabling coherent mode conversion and bandwidth conversion.

\section{Experimental Setup}

Our experimental platform is based on a warm atomic vapor Raman memory. The storage medium is a 75-mm-long cesium-133 vapor cell stabilized at $105\pm1$~$^\circ$C, yielding an effective optical depth of $d=(4.8\pm0.1)\times10^3$. The system utilizes a $\Lambda$-type energy level configuration on the cesium D2 line, with the hyperfine ground states $|F=4\rangle$ and $|F=3\rangle$ serving as the initial state $|\text{g}\rangle$ and the storage state $|\text{s}\rangle$, respectively. Prior to interaction, the atomic ensemble is initialized via optical pumping with an efficiency $(99.6\pm0.1)\%$. 

The coherent signal and control fields are derived from a single continuous-wave laser, with the signal frequency set to 351.7030 THz. To suppress noise arising from four-wave mixing, the memory operates with a large single-photon detuning of $\Delta$ = 18.4 GHz~\cite{Thomas_2019}, corresponding to twice the ground state hyperfine splitting. Both fields are dynamically shaped using an electro-optical modulator driven by arbitrary waveform generators. This configuration enables the carving of complex temporal envelopes and phase profiles required for programmable temporal mode processing. In a typical sequence, a write control pulse maps an input signal (typically shaped with a 50~ns full width at half maximum (FWHM) HG envelope) onto the atomic spin-wave. Following the storage, the signal is retrieved after about 200~ns and is isolated from the strong control fields via a cascaded polarization and spectral filtering stage and detected by superconducting nanowire single-photon detectors. A more detailed description of the experimental setup and pulse generation is included in Appendix A.

For statistical analysis, we construct the dataset by sampling 50 subsets, each containing $1\times10^5$ input-signal detected counts. From these datasets, we calculate the efficiencies and their associated uncertainties. The storage efficiency is determined by comparing the integrated counts of the reference input $N_{\text{ref}}$ and the leaked signal $N_{\text{leak}}$, defined as $\eta_{\text{stor}} = (N_{\text{ref}} - N_{\text{leak}})/N_{\text{ref}}$. The total efficiency is calculated as the ratio of the integrated retrieved signal counts $N_{\text{ret}}$ to the reference input, $\eta_{\text{tot}} = N_{\text{ret}}/N_{\text{ref}}$. All photon counts are background-subtracted prior to calculation.

\section{Temporal mode storage and filtering}

\begin{figure*}
\includegraphics[width=\linewidth]{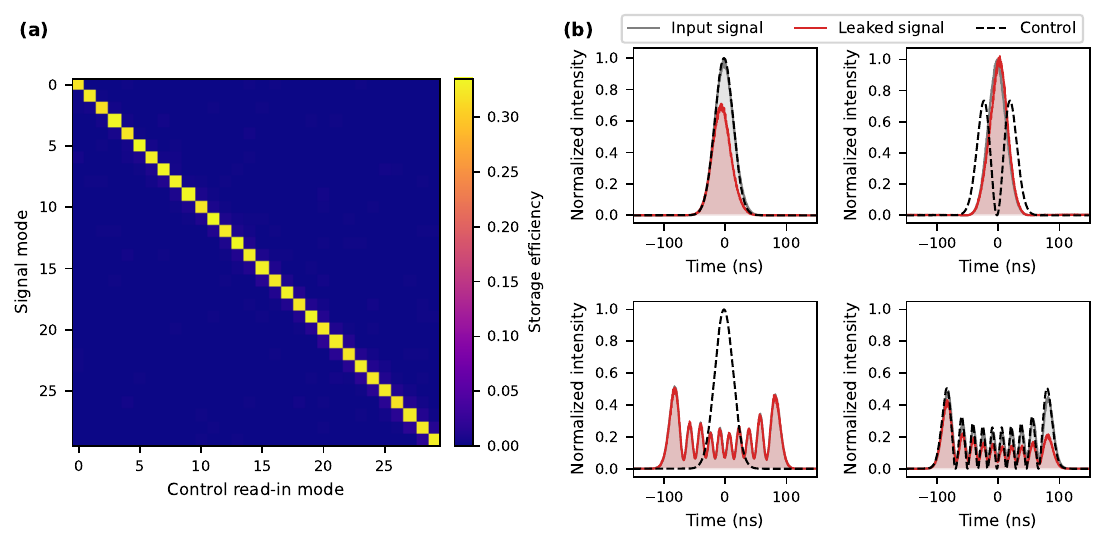}
\caption{\textbf{Temporal mode storage and mode crosstalk.} (a) Crosstalk matrix for the first 30 Hermite-Gaussian temporal modes, represented by the storage efficiencies of different signal-control combinations. Rows correspond to input signal modes, and columns represent control write modes. (b) Examples of storage for different mode combinations. When the control (dashed black) is in the same mode as the input signal (gray), the difference between the input and the leaked signal (red) indicates successful storage. Conversely, when the control and signal modes are orthogonal, minimal storage occurs.}
\label{fig:Crosstalk}
\end{figure*}

\begin{table*}
\caption{\label{tab:table1}%
Examples of component filtering of temporal mode superpositions.
}
\begin{ruledtabular}
\begin{tabular}{lccc}
\textrm{Signal}&
\textrm{Control}&
\textrm{Measured storage efficiency ratio}&
\textrm{Ideal ratio}\\
\colrule
$(\text{HG}_\text{1}+\text{HG}_\text{3})/\sqrt{2}$ & $\text{HG}_\text{1},\text{HG}_\text{3}$ & 0.162(4):0.165(4) & 1:1\\
$(\text{HG}_\text{1}+2\text{HG}_\text{3})/\sqrt{5}$ & $\text{HG}_\text{1},\text{HG}_\text{3}$ & 0.067(4):0.284(4) & 1:4\\
$(\text{HG}_\text{1}+\text{HG}_\text{29})/\sqrt{2}$ & $\text{HG}_\text{1},\text{HG}_\text{29}$ & 0.169(5):0.188(4) & 1:1\\
$(\text{HG}_\text{26}+\text{HG}_\text{28})/\sqrt{2}$ & $\text{HG}_\text{26},\text{HG}_\text{28}$ & 0.158(3):0.157(3) & 1:1\\
$(\text{HG}_\text{1}+\text{HG}_\text{3}$+HG$_\text{5})/\sqrt{3}$ & $\text{HG}_\text{1},\text{HG}_\text{3}$,HG$_\text{5}$ & 0.118(4):0.104(3):0.116(3) &1:1:1\\
\end{tabular}
\end{ruledtabular}
\end{table*}

To establish the Raman memory as a high-dimensional temporal mode-selective buffer, we first benchmarked the storage interaction using a basis set of 30 distinct HG modes ($n=0$ to $29$). This experiment verifies the device's capability to distinguish and store specific temporal waveforms from a dense, high-dimensional alphabet. 

We first probed the memory by preparing weak signal pulses in mode $n$ and applying write control pulses of mode $m$. Consistent with the theoretical kernel formalism, we expect efficient storage only when the signal and control temporal profiles are matched ($n=m$). Using control pulse energies of $128 \pm 1$ $\mu$J, we achieved an average storage efficiency of approximately 30\% for matched modes. The full characterization of this selectivity is presented in the crosstalk matrix in Fig.~\ref{fig:Crosstalk}. The diagonal elements, representing matched signal and control modes, exhibit consistent storage efficiencies ranging from 31.7\% to 33.5\% (with all matrix elements having uncertainties below 0.7\%). Crucially, the off-diagonal elements, representing the crosstalk or leakage of unmatched signal modes, are heavily suppressed. The residual non-zero coupling arises from an interplay of fundamental and technical limits: a fundamental, efficiency-dependent crosstalk intrinsic to the Raman interaction at finite optical depths~\cite{Nunn_2008}, superimposed with imperfections in temporal mode generation. Specifically, we observe a mode-dependent rise in crosstalk as $n$ increases, ranging from $<0.1\%$ to $1.7\%$. We attribute this to the limited bandwidth of our RF generation system, which degrades the fidelity of the rapid temporal features inherent to higher-order modes. The demonstrated dimension of $d=30$ is currently limited by practical constraints, including the optical delay line required to overlap the signal and control pulses and the bandwidth of the RF generation. In principle, however, the ultimate limit is set by the atomic hyperfine splitting, which determines the maximum achievable bandwidth.

Having characterized the basis states, we next investigated the storage of coherent superposition states to demonstrate the memory's ability to selectively filter specific components from a composite signal. We prepared signal pulses in various superpositions of HG modes and applied control fields corresponding to individual component modes. The results, summarized in Table~\ref{tab:table1}, show that the storage efficiency for a given component scales with the square of its probability amplitude, strictly following the projection postulate. The measured storage ratios agree well with the expected weights, confirming that the device functions as a coherent mode filter.

\begin{figure*}
\includegraphics[width=\linewidth]{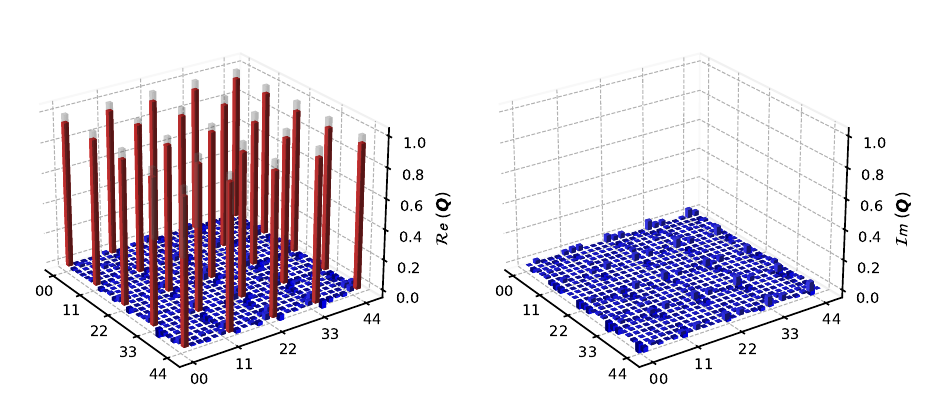}
\caption{\textbf{Storage process matrix.} Real and imaginary components of the reconstructed, efficiency-normalized storage process matrix $\boldsymbol{Q}$. The gray bars represent the ideal process matrix corresponding to a perfect single-mode temporal filter, which is purely real.}
\label{fig:5D_Process_Matrix_3d}
\end{figure*}

Finally, to rigorously certify the filtering capabilities of the storage process, we performed a complete 5-dimensional quantum process tomography within the subspace spanned by the first five HG modes. The storage interaction in Eq.~\ref{eq:storage} can be represented by a linear map $\boldsymbol{\Lambda}$ transforming the input optical field $\mathcal{E}$ and control field $\Omega$ into a stored spin-wave $\mathcal{B}$, as $\mathcal{B} = \boldsymbol{\Lambda}\left(\mathcal{E}\otimes\Omega^*\right)$. Since the spin-wave cannot be directly measured, we characterize the interaction via the storage efficiency $\eta_{\text{stor}}$ similar to Eq.~\ref{eq:storage_eff}: $\eta_{\text{stor}} = (\mathcal{E}\otimes\Omega^*)^\dag \boldsymbol{P} (\mathcal{E}\otimes\Omega^*)$, where $\boldsymbol{P} = \boldsymbol{\Lambda}^\dagger\boldsymbol{\Lambda}$ is the process matrix, which fully encapsulates the storage dynamics.

We probed the system using combinations of signal and control write pulses prepared in mutually unbiased bases and measured the corresponding storage efficiencies (see Appendix B for methods and raw data). Using maximum likelihood estimation under the constraint of complete positivity and trace preservation, we reconstructed the process matrix $\boldsymbol{P}$ of the storage process. To benchmark performance against an ideal single-mode filter, we normalized the result by the storage efficiency to yield $\boldsymbol{Q} = \boldsymbol{P}/\eta_{stor}$. The real and imaginary components of the reconstructed process matrix $\boldsymbol{Q}$ are displayed in Fig.~\ref{fig:5D_Process_Matrix_3d}. In the ideal limit, the matrix element corresponding to matched signal and control modes (e.g. 00, 11, ...) is unity, while all other elements are zeros. Our reconstructed process matrix yields a fidelity of $\mathcal{F} = 0.943 \pm 0.015$ relative to the ideal filter. Furthermore, we evaluated the ``single-modeness" of the interaction, quantified by the ratio:
\begin{equation}    
    \kappa_\Omega = \frac{\max_i \sigma_i}{\text{Tr}(\boldsymbol{P}_\Omega)},
\label{eq:single_modeness}
\end{equation}
where $\{\sigma_i\}$ are the eigenvalues of the reduced process matrix $\boldsymbol{P}_\Omega$, representing the effective quantum channel acting on the signal subspace for a fixed control field $\Omega$. This ratio quantifies the rank of the storage interaction by measuring the fraction of the process weight contained within the dominant eigenmode. Averaging over an ensemble of 1000 randomly sampled control fields, we determined the mean single-modeness to be $0.956 \pm 0.017$. Collectively, these metrics confirm that the Raman storage interaction operates as a near-ideal quantum filter achieving high-fidelity selection of the target temporal mode.



\section{Temporal mode conversion}

The programmability of the Raman memory also enables deterministic manipulation of the stored quantum state. By modifying the temporal profile of the read control field, the stored spin-wave can be retrieved into an optical temporal mode distinct from the input, enabling coherent mode conversion.

\begin{figure*}
\includegraphics[width=\linewidth]{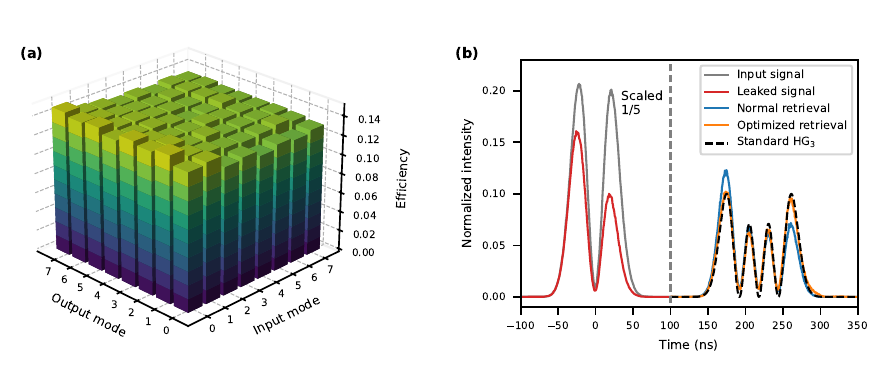}
\caption{\textbf{Mode conversion and optimized retrieval.} (a) Total efficiency for different combinations of mode conversions within the first 8 Hermite-Gaussian modes. (b) Example of retrieval mode optimization converting an HG$_1$ input to an HG$_3$ output. The non-optimized retrieval is shown in blue, and the optimized retrieval is shown in orange. A standard HG$_3$ intensity profile is included for comparison. The input signal (gray) and corresponding leaked signal (red) are scaled by a factor of 1/5.}
\label{fig:mode_conversion}
\end{figure*}

To demonstrate this capability, we performed storage and retrieval experiments across the complete subspace of the first eight HG modes. A signal input in mode HG$_n$ was stored using a matched control field and subsequently retrieved into a target mode HG$_m$ using a reshaped control read pulse. Figure~\ref{fig:mode_conversion}(a) summarizes the internal conversion efficiency for these processes. Using a fixed control pulse energy of $134\pm1$~pJ and a retrieval delay of 218~ns, we observed that the conversion efficiency remains uniform across the mode subspace, ranging from 0.126 to 0.149 with uncertainties below 0.001. This uniformity confirms that the memory's storage and retrieval dynamics are independent of the specific mode index. Fig.~\ref{fig:mode_conversion}(b) details the specific conversion of an input HG$_1$ mode to an output HG$_3$ mode. Using a standard HG$_3$ control pulse for retrieval results in a distorted output (blue trace) characterized by an exponential decay trend superimposed on the pulse structure. This distortion arises from the dynamic depletion of the stored spin-wave; as the excitation is mapped back into the optical field, the retrieval probability is highest at the onset of the control pulse and diminishes as the spin-wave population is exhausted. To compensate for this effect, we engineered the temporal profile of the read control pulse. By applying an exponential growth envelope to the control field while maintaining the same total pulse energy, we counteracted the depletion rate to effectively flatten the retrieval efficiency over the pulse duration. As shown by the orange trace in Fig.~\ref{fig:mode_conversion}(b), this optimized control field successfully retrieves a high-fidelity HG$_3$ mode, demonstrating the system's capacity for precise temporal mode engineering. Additional examples of temporal mode conversion between Gaussian and exponential-decay profiles are detailed in Appendix C.

\begin{figure*}
\includegraphics[width=\linewidth]{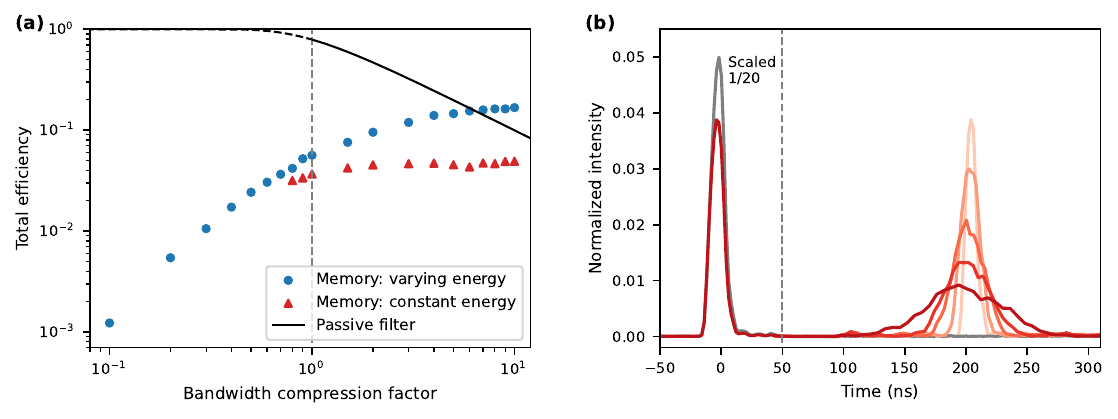}
\caption{\textbf{Bandwidth conversion.} (a) Conversion efficiency of the Raman memory as a function of the bandwidth-compression factor (larger factors correspond to longer output durations). The input signal duration is 10~ns. Blue and red points denote read control pulses with constant peak amplitude and constant energy, respectively. The solid black line shows the response of an ideal passive top-hat frequency filter. Error bars are smaller than the markers. (b) Temporal traces demonstrating conversion using constant-energy control pulses. The 10~ns input yields output durations of 8, 20, 40, 60, and 100~ns (light to dark red). The input signal (gray) and corresponding leaked signal (dark red) are scaled by a factor of 1/20.}
\label{fig:bandwidth_conversion}
\end{figure*}

Beyond reshaping orthogonal mode profiles, the Raman memory facilitates bandwidth conversion by retrieving the stored spin-wave into the same mode profile but with a varied pulse duration. This capability is critical for quantum networking, serving as a bridge between narrowband atomic systems (MHz regime) and high-speed quantum communication channels (GHz regime). We demonstrated this interface by storing a Gaussian signal pulse (HG$_0$) with an electric field FWHM duration of 10~ns. By varying the duration of the retrieval control pulse from 1~ns to 100~ns, we achieved both bandwidth compression and expansion by a factor of 10. We incorporated a booster optical amplifier to expand the pulse energy range required for bandwidth conversion.

The internal total efficiency of these conversion processes is displayed in Fig.~\ref{fig:bandwidth_conversion}(a), where we investigated two distinct retrieval regimes. First, we maintained a constant peak Rabi frequency for the retrieval control field (blue circles). In this regime, the integrated pulse energy scales with duration; consequently, the efficiency increases for longer pulses (bandwidth compression) as the total interaction strength grows, while dropping for shorter pulses (bandwidth expansion) due to reduced control energy. Second, we normalized the control fields to maintain a constant pulse energy ($72\pm2$~pJ) across all durations (red triangles). Representative experimental traces for this regime are shown in Fig.~\ref{fig:bandwidth_conversion}(b). Consistent with Raman interaction theory, where efficiency depends primarily on the integrated control power rather than instantaneous intensity~\cite{Gorshkov_2007}, the conversion efficiency remains relatively constant across a wide range of bandwidth compression factors. In the high-bandwidth expansion regime (short pulses), performance is primarily bounded by the saturation limits of our optical amplification system, which restricts the generation of the necessary high instantaneous Rabi frequencies. With higher power amplification, the system could maintain high internal efficiencies over an even broader conversion range~\cite{Thomas2019}. To benchmark the device's performance, we compared the memory efficiency against the theoretical limit of a passive spectral filter (solid black line). While a passive top-hat filter can achieve bandwidth compression by discarding spectral components, its efficiency scales inversely with the compression factor, resulting in rapid signal loss. Furthermore, passive linear optics are fundamentally incapable of bandwidth expansion (temporal compression). The Raman memory outperforms the passive limit at large compression factors and uniquely enables bidirectional conversion, certifying its capability as an active bandwidth converter.

\section{Discussions}

In this work, we have established the Raman quantum memory as a versatile temporal mode processor. By characterizing the interaction across a high-dimensional Hilbert space, we demonstrated that a single programmable device seamlessly integrates on-demand storage, high-fidelity coherent filtering, deterministic mode conversion, and bidirectional bandwidth conversion. Crucially, operation across the MHz–GHz regime positions the memory as a versatile quantum interface, bridging the spectral gap between narrowband atomic processing nodes and high-speed photonic interconnects. These results confirm that the Raman memory is highly capable of processing complex photonic quantum information carriers, marking it as an essential, active building block for scalable quantum networks.

While the Raman memory offers versatility in temporal mode processing, our current demonstration is limited by relatively low storage efficiencies. Although near-unity efficiencies have been achieved in Raman systems~\cite{Guo_2025}, realizing high-fidelity temporal mode processing is fundamentally constrained by the trade-off between coupling strength and single-modeness. The memory's operation as an ideal single-mode filter relies on the separability of the storage kernel, a condition that strictly holds only in the low-coupling limit. As the optical depth or control power is increased to boost efficiency, the interaction opens coupling channels to higher-order eigenmodes, degrading the mode selectivity~\cite{Nunn_2008,Burdekin_2025}, a behavior confirmed by our numerical simulations (see Appendix D). However, several strategies exist to overcome this limitation. Optimal control techniques can tailor the control pulse shapes to maximize efficiency for a target mode while actively suppressing coupling to orthogonal eigenmodes~\cite{Gorshkov_2008,Guo_2025}. Alternatively, structural modifications such as the Efficiency Enhancement via Light-Matter Interference (EEVI) protocol offer a path to high efficiency at lower single-pass coupling strengths~\cite{Burdekin_2025}. By exploiting interference within a loop geometry, EEVI distributes the interaction across multiple stages, allowing reduced control power per pass to preserve single-modeness while cumulatively achieving high efficiency. Cavity-enhanced memories provide a comparable solution~\cite{Nunn_2017,Saunders_2016}, though they are typically bandwidth-constrained by the cavity finesse. Implementing these protocols would improve efficiency while preserving mode selectivity.

The realization of this high-dimensional temporal mode processor opens several promising avenues in quantum information science. First, in high-dimensional quantum key distribution~\cite{Cozzolino_2019,Ogrodnik_2025}, encoding in a $d$-dimensional temporal basis increases channel capacity by $\log_2(d)$ and enhances noise resilience. The memory's ability to directly filter and measure these modes in the time domain provides a scalable receiver architecture for such networks. Second, its combined capacity for coherent filtering, mode reshaping, and bandwidth conversion establishes the processor as a critical interconnect between different nodes in networks~\cite{Hammerer_2010,Gao_2019}. It acts as an optimal coherent filter for solid-state emitters, buffering the dominant eigenmode to enhance photon indistinguishability without the severe brightness penalties of passive filtering. By programming the readout field, the memory unifies modes from disparate, noisy sources into identical photons and bridges the spectral gap between broadband photonic channels and narrowband atomic nodes. Finally, the high-fidelity temporal mode filtering of our platform enables optimal measurements for quantum-enhanced metrology in the time-frequency domain~\cite{Mazelanik_2022,Ansari_2021,Zhang_2025}, facilitating super-resolved parameter estimation tasks like time-of-flight ranging and spectroscopy. Extending this technique to high-dimensional temporal mode operation further unlocks multiparameter estimation capabilities. Ultimately, these integrated capabilities elevate the Raman platform to an active, versatile node for the future hybrid quantum internet.

\begin{acknowledgments}
We would like to acknowledge Benjamin Brecht and Joseph H. D. Munns for useful contributions in the early stages of this project. We thank Micha{\l} Karpi\'nski, Raj B. Patel, and Zhenghao Li for helpful discussions. This work was supported by the European Union’s Horizon 2020 Research and Innovation Programme Grant No. 899587 Stormytune and the Engineering and Physical Sciences Research Council via the Quantum Computing and Simulation Hub (Grant No. T001062). A. Z. acknowledges a UK Research and Innovation Guarantee Postdoctoral Fellowship under the UK government’s Horizon Europe funding Guarantee (EP/Y029127/1). S.E.T. acknowledges an Imperial College Research Fellowship.
\end{acknowledgments}


\appendix

\section{Detailed Experimental Setup}

\begin{figure*}
\includegraphics[width=0.7\linewidth]{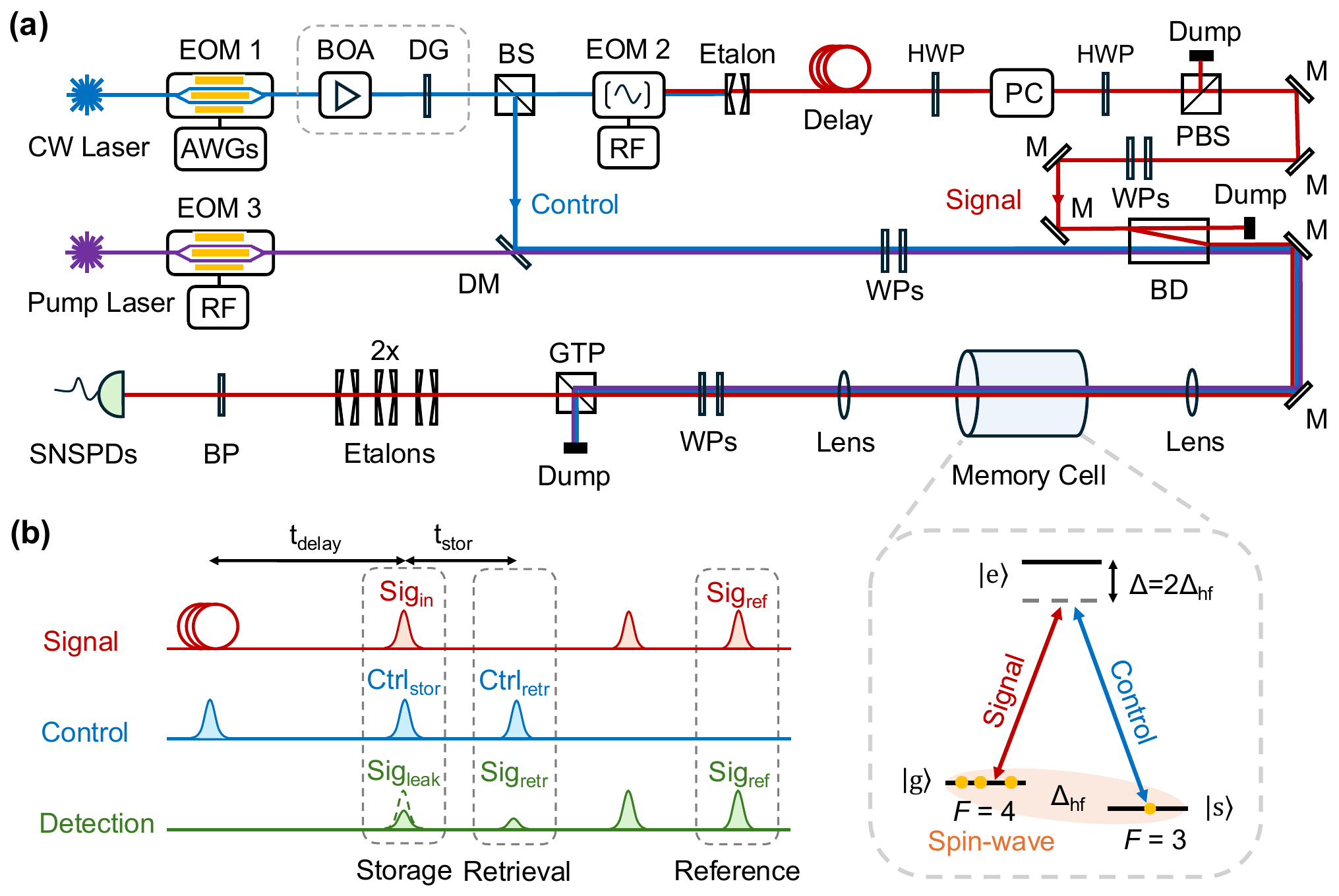}
\caption{\textbf{Experimental setup}. (a) Schematic of the Raman memory apparatus. Signal and control fields are derived from a single continuous-wave (CW) laser and carved into pulses by an electro-optical modulator (EOM 1). A beam splitter (BS) separates the paths; the signal field is frequency-shifted by EOM 2, isolated by an etalon, and passed through a fiber delay line for synchronization. The fields are spatially combined using a beam displacer (BD) before interacting in the memory cell. Post-interaction filtering is performed by a Glan-Taylor polarizer (GTP), Fabry-Pérot etalons, and a band-pass filter (BP) prior to detection by superconducting nanowire single-photon detectors (SNSPDs). (AWGs: arbitrary waveform generators; BOA: booster optical amplifier; DG: diffraction gratings; HWP: half-wave plate; PC: Pockels cell; PBS: polarizing beam-splitter; M: mirror; WPs: a quarter-wave plate and a half-wave plate; DM: dichroic mirror). The inset depicts the $^{133}$Cs $\Lambda$-level scheme, where signal and control fields couple the hyperfine ground states $|1\rangle$ and $|3\rangle$ with a detuning $\Delta = 2\Delta_{\text{hf}}$ from the excited state $|2\rangle$. (b) Pulse sequence. The timing diagram illustrates the storage of an input signal ($\text{Sig}_{\text{in}}$) via a write control pulse ($\text{Ctrl}_{\text{stor}}$) and subsequent retrieval ($\text{Sig}_{\text{retr}}$) by a read pulse ($\text{Ctrl}_{\text{retr}}$).}
\label{fig:setup}
\end{figure*}


The experimental apparatus, illustrated in Fig.~\ref{fig:setup}(a), comprises the complete setup of the warm atomic vapor Raman memory. The cesium vapor cell is enclosed within a three-layer $\mu$-metal magnetic shield to suppress external magnetic field fluctuations. The atomic ensemble is initialized in the ground state $|g\rangle$ via optical pumping on the D1 transition. Coherent optical fields are derived from a single continuous-wave external cavity diode laser (Toptica DL Pro) tuned to the control frequency of 351.7122~THz. The laser output is modulated by an intensity-phase electro-optical modulator (EOM 1, Sacher Lasertechnik AM830PF) driven by two arbitrary waveform generators (Tektronix AWG70001A), allowing for the carving of precise temporal pulse shapes with complex phase profiles. (For bandwidth conversion experiments requiring high Rabi frequencies, we amplified the pulses with a booster optical amplifier followed by a diffraction grating to suppress amplified spontaneous emission.) A beam splitter directs 90\% of the optical power to the control field path. The remaining 10\% is modulated by a second EOM to generate frequency sidebands, from which the signal field at 351.7030~THz is isolated using a Fabry-P\'erot etalon. To ensure temporal synchronization with the control pulses, the signal field traverses a fiber delay line before entering the memory. The signal also passes through a Pockels cell, which is used to suppress background light within the retrieval time window, thereby reducing noise. Finally, the orthogonally polarized signal and control fields are spatially combined on a beam displacer and focused into the vapor cell , with measured beam waists at the interaction center of $164\pm5$~$\mu$m and $190\pm5$~$\mu$m, respectively.

The experimental sequence, illustrated in Fig.~\ref{fig:setup}~(b), consists of a synchronized storage and retrieval operation. A write control pulse maps the input signal pulse onto a collective atomic spin-wave excitation; following a programmable delay, a read control pulse retrieves the excitation back into the optical domain. Except for bandwidth conversion tasks, all optical pulses are shaped with a Hermite-Gaussian temporal envelope with an electric field full width at half maximum (FWHM) of 50~ns. 

Following the memory interaction, the strong control and pump fields are filtered from the retrieved signal. This is accomplished via a cascaded filtering stage consisting of a Glan-Taylor polarizer, three double-passed Fabry-P\'erot etalons, and a narrow bandpass filter. While this configuration achieves a suppression factor of $4\times10^7$, it limits the total signal transmission efficiency to approximately 3\%. The filtered signals are detected by superconducting nanowire single-photon detectors (Photon Spot), and arrival times are recorded by a time tagger (Swabian TimeTagger20). 


The high-fidelity manipulation of high-dimensional temporal modes relies critically on the precise synthesis of the optical signal and control fields in both amplitude and phase. Here, we detail the generation scheme and the methods used to correct for electronic distortions.

We employ a single intensity-phase electro-optical modulator driven by two independent arbitrary waveform generators and RF amplifiers (Tektronix PSPL5868). This dual-drive configuration allows for independent control over the phase introduced by each waveguide arm. To ensure a high extinction ratio, a bias voltage $V_{\text{bias}}(t)$ is actively feedback-controlled to maintain the modulator at its destructive interference (null) point in the absence of drive signals. In this configuration, the output field's normalized amplitude $A(t)$ and global phase $\Phi(t)$ are related to the drive voltages $V_1(t)$ and $V_2(t)$ by:
\begin{equation}
\begin{aligned}
A(t)    &= \sin\left(\frac{\pi [V_1(t)+V_2(t)]}{2V_\pi}\right), \\
\Phi(t) &= \frac{\pi [V_1(t)-V_2(t)]}{2V_\pi}.
\end{aligned}
\label{eq:EOM_relations}
\end{equation}
Here, $V_\pi$ is the half-wave voltage, defined as the voltage required to induce a $\pi$ phase shift in a single waveguide arm. To generate a target pulse shape defined by $A(t)$ and $\Phi(t)$, the required input voltages are derived by inverting these relations:
\begin{equation}
\begin{aligned}
V_1(t) &= \frac{V_\pi}{\pi}\left[\arcsin(A(t))+\Phi(t)\right], \\
V_2(t) &= \frac{V_\pi}{\pi}\left[\arcsin(A(t))-\Phi(t)\right].
\end{aligned}
\label{eq:EOM_drive}
\end{equation}

To achieve high generation fidelity, we implemented specific correction protocols:

1. Bias Voltage Stabilization: Precise bias control is essential for defining the null point. Conventional minimization algorithms, which dither the bias to minimize intensity, fail for high-order HG modes where the temporal intensity profile is oscillating and non-zero. To resolve this, bias calibration is strictly performed without RF inputs ($V_1=V_2=0$) to establish the correct extinction point before generating higher-order modes.

2. Frequency Response Correction: Imperfections in the RF signal chain (AWGs, amplifiers, cables) distort the driving waveforms. We approximate the electronic system as a linear time-invariant system characterized by a frequency response function $\tilde{R}(\omega) = \tilde{h}(\omega)/\tilde{x}(\omega)$, where $\tilde{h}$ and $\tilde{x}$ are the Fourier transforms of the output and input signals, respectively. By characterizing $\tilde{R}(\omega)$ using a broadband Gaussian probe pulse, we pre-distort the input waveform $x'(t)$ to generate the target shape $h'(t)$:\begin{equation}x'(t) = \mathcal{F}^{-1}\left[\frac{\mathcal{F}(h'(t))}{\tilde{R}(\omega)}\right].\label{eq:correction}\end{equation}

3. Electrode Crosstalk Compensation: We observed a discrepancy in the effective $V_\pi$ required for intensity versus phase modulation, attributed to capacitive crosstalk between the RF electrodes. We model this as a linear mixing where the effective voltage on one arm is perturbed by the other (e.g. $V_{1,\text{eff}} = V_1 + \epsilon V_2$). When these effective voltages are substituted into Eq.~\ref{eq:EOM_drive}, the predicted effective $V_\pi$ agreed with our experimental measurements. We thus compensated for this coupling by adjusting the phase term in Eq.~\ref{eq:EOM_drive} when generating the pulses, ensuring simultaneous high-fidelity control of both amplitude and phase.

4. Low-Frequency Cutoff Compensation: Commercial RF amplifiers typically exhibit a low-frequency cutoff, effectively blocking the DC component of the signal. This introduces a vertical voltage shift that distorts the generated optical envelope. This effect is particularly acute for even-order HG modes, which inherently possess significant spectral weight at low frequencies, leading to imperfect pulse shaping for these specific modes. We compensate for this by applying a small DC voltage to the RF amplifiers.

5. Energy Correction: Non-linearities in the RF amplification chain can result in mode-dependent energy variations. To address this, we measured the optical pulse energies for the full basis set and normalized the digital waveforms relative to the HG$_0$ energy. This active pre-compensation successfully reduced energy variations to $<1\%$.

We note that our current pulse-generation scheme, which relies on a single intensity-phase EOM, is constrained by the limited dynamic range of the RF drivers. This gives abrupt modulo-$2\pi$ phase resets to restrict the phase evolution. These sharp discontinuities require infinite RF bandwidth to reproduce, resulting in imperfect pulse shapes and phase fidelity. Transitioning to IQ (In-Phase/Quadrature) modulation~\cite{Xu_2020} would resolve this limitation by enabling the independent and continuous modulation of the real and imaginary field components, thereby significantly improving the generation fidelity of complex high-dimensional states.



\section{Quantum Process Tomography and MUB Generation}

\begin{figure*}
\includegraphics[width=0.8\linewidth]{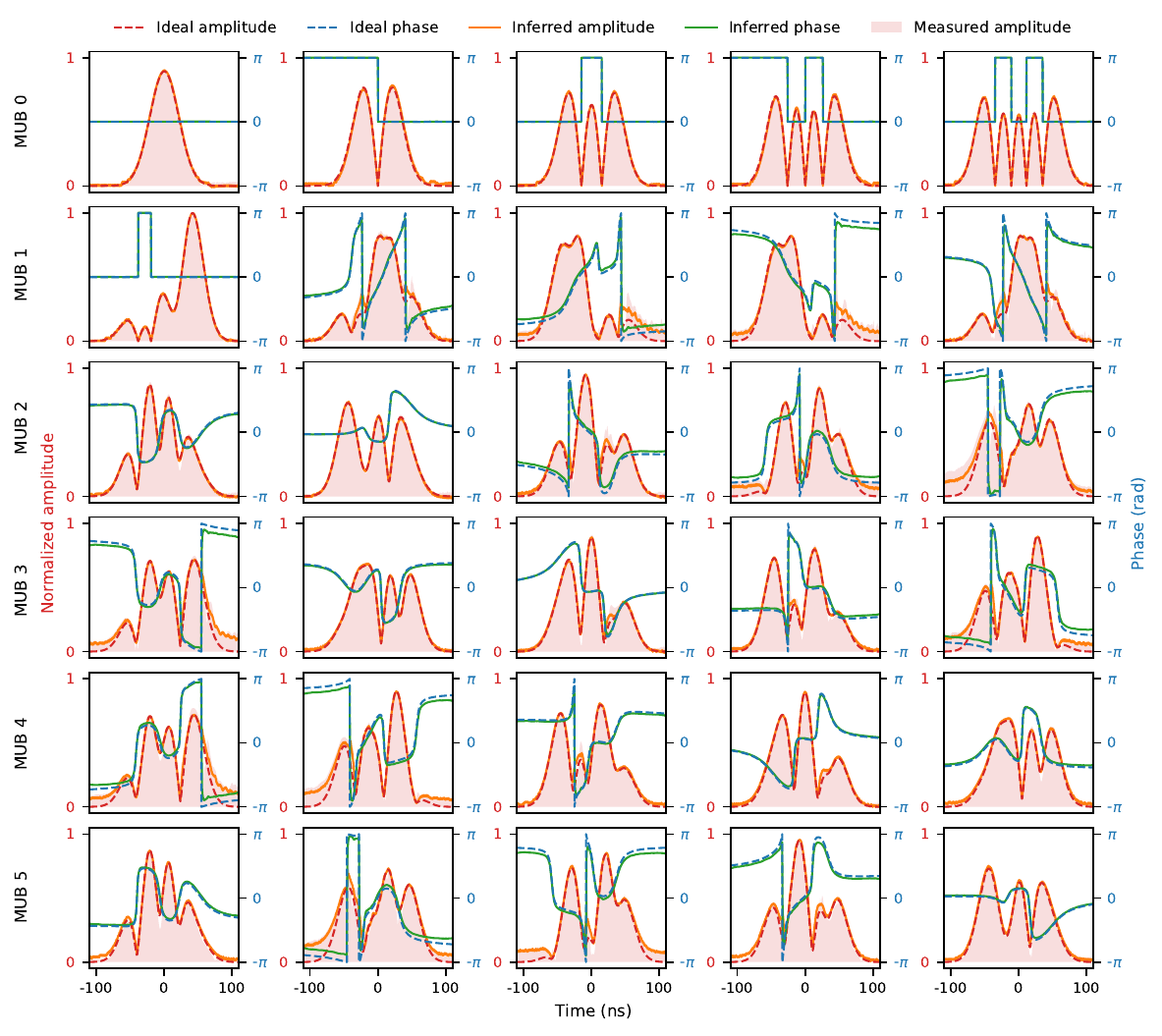}
\caption{\textbf{MUB state generation.} Mutually unbiased basis states generated for 5D process tomography. Dashed blue and red lines represent the ideal amplitude and phase, respectively. Solid orange and green lines show the amplitude and phase inferred from the measured RF signals applied to the EOM. The shaded red area indicates the amplitude derived from intensity measurements taken with a photodiode.}
\label{fig:sup_MUBs}
\end{figure*}

To fully reconstruct the storage process dynamics, the system must be probed using a tomographically complete set of states. We utilize Mutually Unbiased Bases (MUBs) constructed within the $d=5$ dimensional subspace spanned by the first five Hermite-Gaussian modes~\cite{Durt_2010,Ansari_2017}. This basis choice ensures that both the amplitude and phase response of the memory are rigorously sampled. For a Hilbert space of dimension $d$, there exist $d+1$ sets of MUBs, each containing $d$ orthogonal basis states. These include the computational basis (the HG eigenstates $\left\{|0\rangle,|1\rangle,...,|d-1\rangle\right\}$) and $d$ bases formed by linear superpositions with equal magnitudes but distinct phase relations (Fourier bases). The experimental realization of these states is visualized in Fig.~\ref{fig:sup_MUBs}. The plots compare the ideal target waveforms (dashed lines) with the measured optical amplitude (red shaded regions) and the complex field inferred from the applied RF drive signals (solid lines), demonstrating good agreement.

\begin{figure*}
\includegraphics[width=0.8\linewidth]{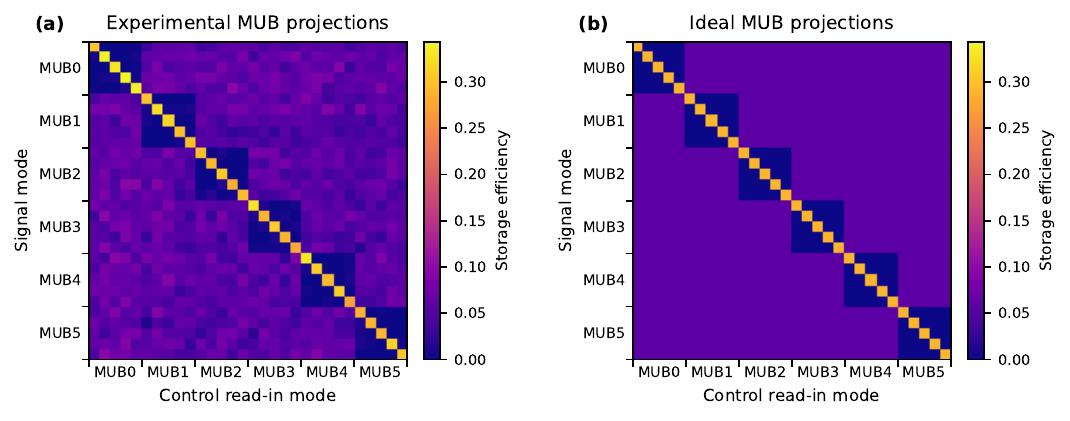}
\caption{\textbf{MUB projections measured as storage efficiencies.} (a) Storage efficiencies measured for different combinations of signal and control modes prepared in MUB states. (b) Ideal storage efficiencies for MUB projections assuming an ideal single-mode memory.}
\label{fig:sup_MUB_projections}
\end{figure*}

By preparing all pairwise combinations of signal and control fields in these MUB states, we probe the full bipartite Hilbert space. The storage process is described by a process matrix $\boldsymbol{P}$, a $d^4$-dimensional Hermitian matrix containing $2d^4$ real parameters. By constraining the process to be Completely Positive and Trace-Preserving —a requirement for any physical quantum channel—the number of independent parameters is reduced to $d^4-d^2$. Our experimental dataset, comprising $(d(d+1))^2$ independent storage efficiency measurements, is significantly overcomplete relative to this parameter count, ensuring a robust and accurate reconstruction via maximum likelihood estimation. The raw storage efficiencies for these MUB projections are displayed in Fig.~\ref{fig:sup_MUB_projections}. The high efficiencies along the diagonal and the strong suppression of off-diagonal elements within each basis set confirm the high-fidelity generation and orthogonality of the MUB states. These measurements were used to reconstruct the process matrix presented in the main text (Fig.~\ref{fig:5D_Process_Matrix_3d}).

Finally, we quantify the single-modeness of the memory for a specific control field $\Omega$ by projecting the full process matrix onto the control subspace: $\boldsymbol{P}_\Omega=(\boldsymbol{I}\otimes\Omega^*)^\dag\boldsymbol{P}(\boldsymbol{I}\otimes\Omega^*)$. As defined in the main text Eq.~\ref{eq:single_modeness}, the single-modeness metric $\kappa_\Omega$ is calculated from the eigenvalues of this reduced matrix $\boldsymbol{P}_\Omega$, representing the purity of the temporal filtering operation.



\section{Mode conversion examples}

\begin{figure*}
\includegraphics[width=0.8\linewidth]{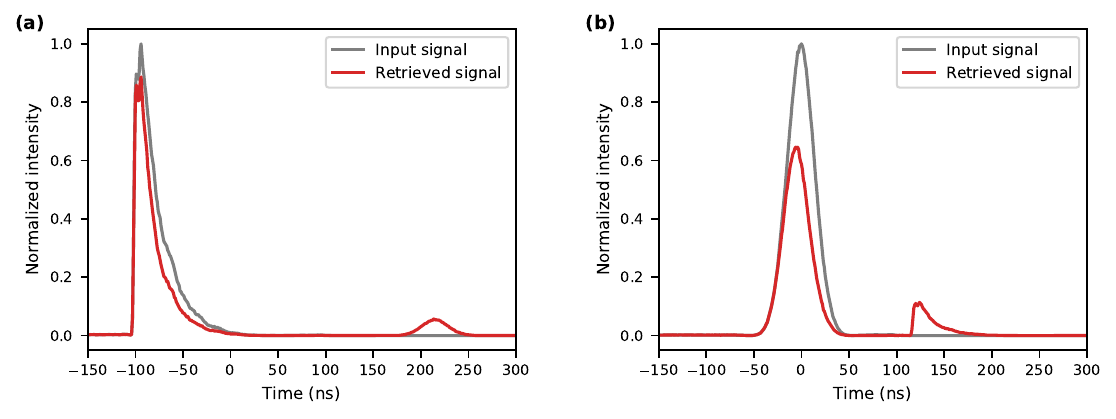}
\caption{\textbf{Temporal mode conversion between Gaussian and exponential-decay profiles.} (a) A signal with an exponentially decaying temporal profile is converted into a Gaussian pulse. (b) A Gaussian signal is converted into an exponentially decaying pulse.}
\label{fig:covert}
\end{figure*}

Figure~\ref{fig:covert} illustrates two examples of temporal mode conversion between Gaussian and exponential-decay pulse shapes. In the first example, an exponentially decaying signal with a 50~ns decay constant is stored using a control pulse with an identical temporal profile, yielding a storage efficiency of $(27.3\pm1.1)\%$. The signal is subsequently retrieved using a Gaussian control pulse, resulting in a total internal efficiency of $(6.5\pm0.3)\%$. Conversely, a Gaussian signal is stored via a Gaussian control pulse with a $(38.9\pm0.9)\%$ storage efficiency, and then retrieved as an exponentially decaying pulse with a total internal efficiency of $(8.0\pm0.2)\%$. Note that in this demonstration, the pulse energy of the exponentially decaying pulse is lower than that of the standard Gaussian pulse. Ultimately, this temporal mode conversion highlights the capability of our Raman memory to dynamically manipulate the temporal profiles of signal pulses, enabling seamless interfacing with photon sources of varying temporal shapes.

\section{Coupling Strength and Control Pulse Optimization}

\begin{figure*}
\includegraphics[width=0.8\linewidth]{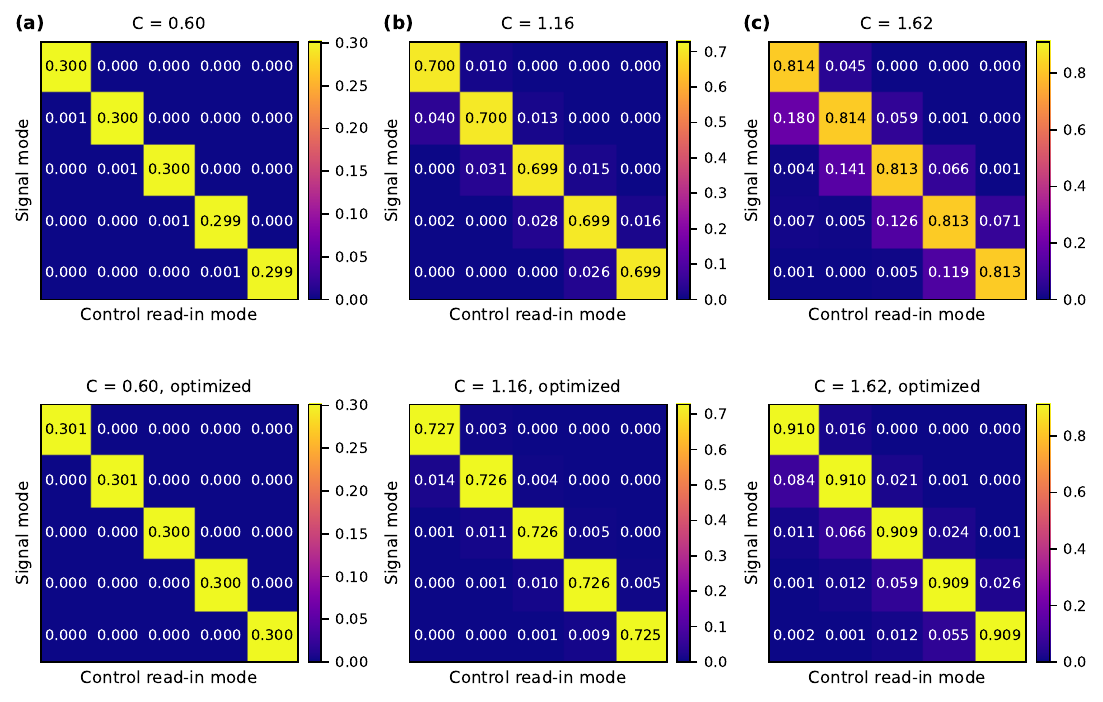}
\caption{\textbf{Simulated mode crosstalk and control optimization.} Simulated Hermite-Gaussian mode crosstalk matrices for the storage process at different coupling strengths, compared to matrices obtained using control write pulses optimized for maximum storage. The coupling strengths are (a) $C=0.60$, (b) $C=1.16$, and (c) $C=1.62$. While crosstalk naturally increases with higher coupling strength, the control optimization effectively suppresses it.}
\label{fig:sup_crosstalk}
\end{figure*}

\begin{figure*}
\includegraphics[width=0.8\linewidth]{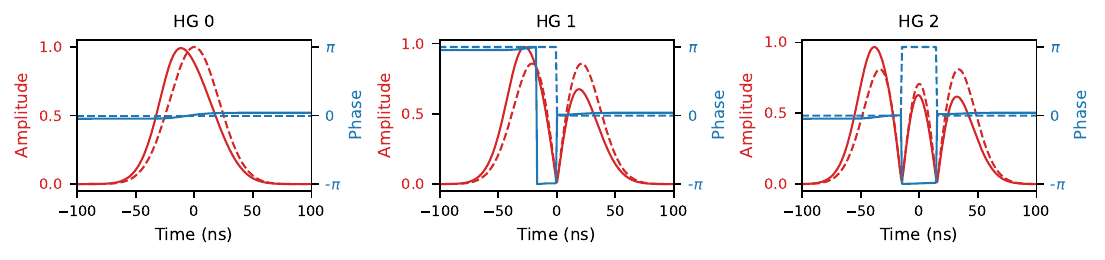}
\caption{\textbf{Simulated control write pulse optimization for the storage process.} The simulated amplitude (red) and phase (blue) are shown for standard Hermite-Gaussian modes 0, 1, and 2 (dashed lines) compared to controls optimized for storage (solid lines), while maintaining the same pulse energy.}
\label{fig:sup_crosstalk_control}
\end{figure*}

\begin{figure*}
\includegraphics[width=0.8\linewidth]{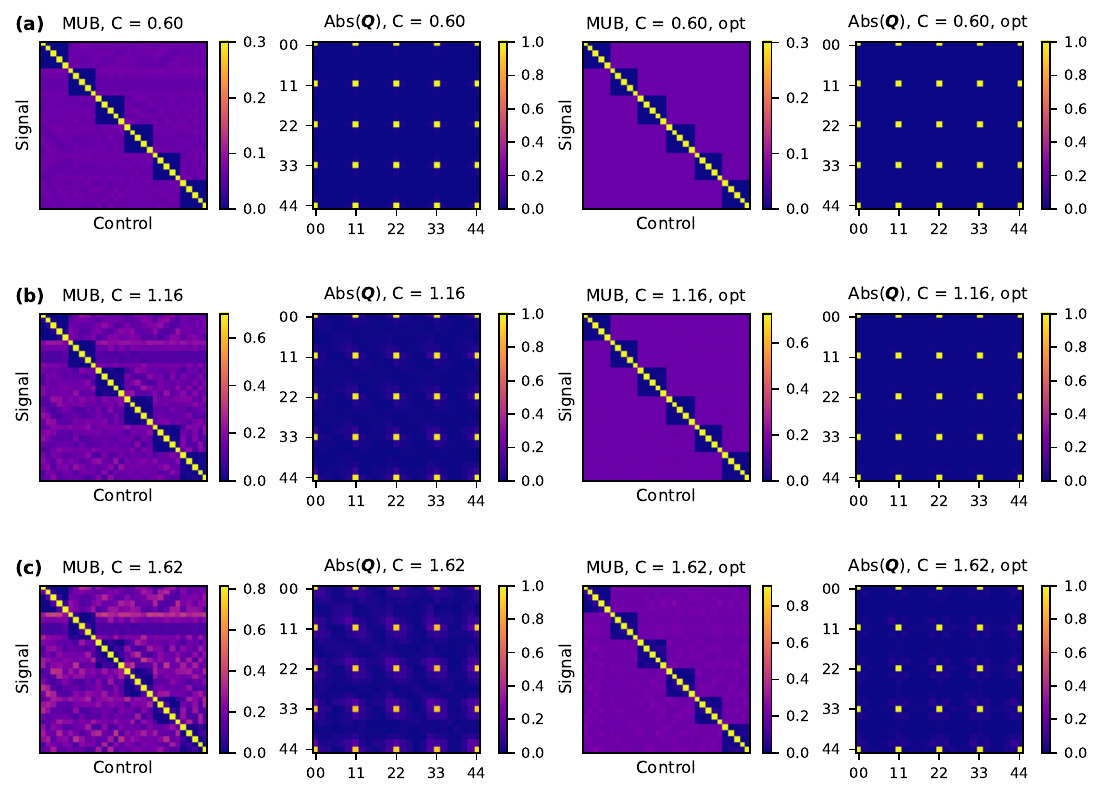}
\caption{\textbf{Simulation of 5D process tomography with varying coupling strengths and control optimization.} Simulated MUB projections and tomography results for the first five Hermite-Gaussian modes. The panels show performance at coupling strengths of (a) $C=0.60$, (b) $C=1.16$, and (c) $C=1.62$. Each panel compares the standard response with the results obtained using optimized control pulses.}
\label{fig:sup_process_tomo}
\end{figure*}

A fundamental trade-off exists in Raman storage between efficiency and mode selectivity. Increasing the coupling strength $C$ (proportional to the square root of optical depth or control pulse energy) enhances storage efficiency but simultaneously degrades single-modeness, as the storage kernel becomes increasingly non-separable. Here, we investigate this relationship via numerical simulations and demonstrate how optimal control theory can mitigate selectivity loss. The simulation follow the methods in Ref.~\cite{Burdekin_2025}. We employ the Krotov method~\cite{Gorshkov_2008} to iteratively find the specific control pulse shape (amplitude and phase) that maximizes storage efficiency for a target signal mode while constraining the total control pulse energy to remain constant.

We first simulated the mode crosstalk within the subspace of the first five HG modes. Figure~\ref{fig:sup_crosstalk} displays the storage efficiency matrices for coupling strengths $C=0.60, 1.16,$ and $1.62$. Using standard HG control pulses, increasing $C$ from $0.60$ to $1.62$ boosts the diagonal storage efficiency from 30.0\% to 81.4\%; however, this is accompanied by a severe rise in nearest-neighbor crosstalk (the first off-diagonal term) from 0.1\% to 18.0\%. By applying control optimization, we significantly improve performance. The optimized controls increase the diagonal efficiencies to 30.1\% and 91.0\%, respectively, while suppressing the off-diagonal crosstalk from negligible levels ($<0.1\%$) to only 8.4\% at the highest coupling strength. Figure~\ref{fig:sup_crosstalk_control} illustrates the optimized control fields for the first three HG modes at $C=1.62$. The optimized waveforms exhibit a characteristic temporal skew toward the leading edge and subtle phase modulations, features which actively compensate for the dynamic variation of the spin-wave distribution during storage.

To quantify the global improvement in quantum performance, we simulated a 5-dimensional quantum process tomography for both standard and optimized controls (Fig.~\ref{fig:sup_process_tomo}). For standard pulses, as $C$ increases from $0.60$ to $1.62$, the process fidelity relative to an ideal single-mode filter degrades from 99.8\% to 90.6\%, and the single-modeness $\kappa_\Omega$ drops from $(99.6\pm0.1)\%$ to $(81.9\pm4.2)\%$. Control pulse optimization effectively recovers the mode selectivity: the fidelities are restored to 99.9\%, 99.7\%, and 98.7\%, while the single-modeness is maintained at $(99.9\pm0.0)\%$, $(98.5\pm0.3)\%$, and $(93.4\pm1.5)\%$, respectively. These results confirm that pulse optimization is a viable route to achieving high-efficiency temporal mode filtering without sacrificing the single-modeness required for quantum information processing.

\bibliography{TF_Filter}



\nocite{*}


\end{document}